\documentclass[aps,pre,twocolumn,showpacs,superscriptaddress]{revtex4-1}

\usepackage[dvipdfmx]{graphicx,color}
\usepackage{amsmath,amssymb,bm,amsthm,amsfonts}
\usepackage{physics}
\usepackage{lipsum}
\usepackage{braket}

\begin{document}

\title{Finite-temperature properties of extended Nagaoka ferromagnetism}

\author{Hiroaki Onishi}
\email{onishi.hiroaki@jaea.go.jp}
\affiliation{%
Advanced Science Research Center,
Japan Atomic Energy Agency,
Tokai, Ibaraki 319-1195, Japan
}
\author{Seiji Miyashita}
\affiliation{%
Department of Physics, Graduate School of Science,
The University of Tokyo,
7-3-1 Hongo, Bunkyo-Ku, Tokyo 113-0033, Japan
}

\date{\today}

\begin{abstract}
We study finite-temperature properties of a Hubbard model including sites of a particle bath
which was proposed as a microscopic model to show itinerant ferromagnetism at finite electron density.
We use direct numerical methods, such as exact diagonalization and random vector methods.
The temperature dependence of quantities is surveyed in the full range of the temperature.
We find that the specific heat has several peaks,
which correspond to ordering processes in different energy scales.
In particular, magnetic order appears at very low temperature.
Depending on the chemical potential of the particle bath and the Coulomb interaction,
the system exhibits an itinerant ferromagnetic state or an antiferromagnetic state of the Mott insulator.
Microscopically the competition between these two types of orderings
causes a peculiar ordering process of local spin correlations.
Some local ferromagnetic correlations are found to be robust,
which indicates that the ferromagnetic correlation originates from the motion of itinerant electrons
in a short-range cluster.
\end{abstract}

\maketitle

\section{Introduction}

It is a long-standing subject to understand magnetic properties and their origins
in solids~\cite{Heisenberg1928,Slater1936,Stoner1938,Slater1953}.
Since Heisenberg pointed out that the exchange energy of electrons is a key ingredient 
to give magnetic interactions~\cite{Heisenberg1928},
various studies have been done for the mechanism of magnetic orderings.
In particular, mechanisms to produce a ferromagnetic (FM) state have been studied extensively.
For localized spin systems, magnetic properties are modeled by 
the Heisenberg model with the exchange coupling of spins on different sites.
The exchange coupling is FM in some situations~\cite{Goodenough1955,Kanamori1957a,Kanamori1957b,Goodenough1958,Kanamori1959}.
Based on the Heisenberg model,
detailed magnetic properties including phase transitions and critical phenomena have been clarified.

On the other hand, for itinerant electron systems,
the motion of electrons plays an important role in magnetic properties.
Stoner gave an essential idea known as the Stoner criterion for the FM ordering
using a mean-field analysis~\cite{Stoner1938}.
In this direction,
first-principles methods based on the density functional theory have been widely used
to study the electronic band structure and magnetic properties~\cite{Kohn1999,Jones2015}.
This type of calculations enables us to explain the ground-state property
such as magnetizations of prototypical FM metals Fe, Co, and Ni~\cite{Janak1979,Akai1990}.
However, by this approach,
it is difficult to properly explain the temperature dependence of thermodynamic quantities,
and to take account of correlations among electrons.
A systematic treatment of collective electron correlations has been developed as
the self-consistent renormalization (SCR) theory of spin fluctuations~\cite{Moriya1973a,Moriya1973b,Moriya-book1985},
and it has succeeded in reproducing the Curie-Weiss law.

Focusing on the many-body effects of electrons,
the Hubbard model~\cite{Hubbard1963,Kanamori1963,Gutzwiller1963}
was introduced,
and properties of the so-called strongly correlated materials have been studied from a microscopic viewpoint.
To date, several itinerant electron models which exhibit FM order in the ground state have been proposed.
In the Hubbard model, no explicit dependence on spin states exists
and the FM order comes out due to the electron motion under the Pauli exclusion principle.
The Nagaoka ferromagnetism is an example in which
the emergence of the FM ground state is rigorously proven in the Hubbard model~\cite{Nagaoka1966,Thouless1965,Tasaki1989}.
The flat-band ferromagnetism is another example~\cite{Mielke1991a,Mielke1991b,Mielke1992,Tasaki1992,Mielke1993,Tasaki1994,Tasaki1996}.
In multi-orbital systems,
the FM order is induced by the Hund's rule coupling~\cite{Kubo1982,Kusakabe1994,Onishi2004,Onishi2007a,Onishi2007b}.
Coupled systems of itinerant electrons and localized spins show the FM spin alignment
by the double exchange mechanism~\cite{Zener1951,Anderson1955,deGennes1960},
described by the Kondo lattice model~\cite{Sigrist1991,Sigrist1992,Tsunetsugu1997,Nolting2003,Yamamoto2010}.
However, the temperature dependence of magnetic properties of itinerant magnets
is still far from being understood
due to the difficulty in accurately calculating properties at finite temperatures.

In this paper, we study a kind of Nagaoka FM model.
The Nagaoka ferromagnetism takes place
in systems with one hole added to the half-filling
and infinitely large Coulomb interaction
on lattices satisfying the so-called connectivity condition.
It indicates that the motion of a hole causes a nodeless wavefunction
that corresponds to a saturated FM state.
In contrast, it is known that, when the system is at half-filling,
the ground state is a Mott state with a charge gap
and an antiferromagnetic (AFM) correlation between neighboring spins.
Thus, the change by one electron gives the striking difference.
It is, however, impossible to control one electron in a bulk system,
and the Nagaoka ferromagnetism is ill-defined in the thermodynamic limit.
Then,
effects of more than one hole and finite Coulomb interaction have been examined~\cite{Takahashi1982,Doucot1989,Riera1989,Fang1989,Shastry1990,Suto1991,Toth1991,Hanisch1993,Hanisch1995,Sakamoto1996,Arita1998,Daul1997,Daul1998,Kohno1997,Watanabe1997a,Watanabe1997b,Watanabe1999}.
It has been shown that in some cases the Nagaoka ferromagnetism is destroyed with more than one hole,
and properties in the thermodynamic limit has not been clear.

To clarify the relation of the Nagaoka ferromagnetism to a macroscopic FM state,
it is important to establish concrete models which show the itinerant ferromagnetism
in a finite range of hole density (not a number) in the thermodynamic limit.
In this context, we have proposed a model
in which the electron density can be controlled continuously
and a FM state is realized in the thermodynamic limit~\cite{Miyashita2008,Onishi2014}.
We study a system on a lattice which consists of two parts:
a part regarded as a main frame
and a part which works as a particle bath.
The electron density in the main frame is controlled
by the chemical potential of the particle bath.
We have reported that a transition between Mott AFM and itinerant FM states really occurs at zero temperature
when varying the difference of the chemical potentials between the main frame and the particle bath.
We call this type of FM state ``extended Nagaoka FM state''.

Here we note that
the mechanism of ferromagnetism in the itinerant electron system
is different from that in the localized spin system.
That is, in the localized spin system,
the FM ordering is caused by local FM exchange interactions.
In contrast, in the itinerant electron system,
there is no explicit FM interaction between spins,
but mobile electrons or holes traveling in the whole system
causes the ferromagnetism.
Therefore, we naively expect some difference in ordering properties
between itinerant electrons and localized spins at finite temperatures.

In this paper,
we study finite-temperature properties of a model for the extended Nagaoka ferromagnetism by numerical methods,
such as exact diagonalization (ED)
and random vector methods~\cite{Hams2000,Jin2021,Sugiura2013}.
We survey ordering processes in the full range of the temperature.
We find that the specific heat has several peaks,
which correspond to ordering processes in different energy scales.
At a high temperature of the order of the Coulomb interaction $U$,
a peak appears due to the suppression of double occupancy.
At a temperature of the order of the electron hopping $t$
and the chemical potential of the particle bath $\mu$,
we find another peak due to the settlement of optimal electron distribution.
At much low temperature,
we find peak(s) due to magnetic ordering.
In the vicinity of a quantum phase transition between FM and AFM states,
we find another peak which resembles the quasi-gap behavior in the high-$T_{\mathrm{c}}$ superconducting systems. 

To study the magnetic orderings microscopically, we investigate spin correlation functions.
We show that the competition between the FM and AFM orderings
causes a peculiar ordering process.
There, local FM correlations in a cluster are found to be robust,
which indicates that the FM correlation originates from the motion of itinerant electrons
in the cluster.

The rest of the paper consists of the following sections:
In Sec.~II, we explain a model for an extended Nagaoka ferromagnetism.
In Sec.~III and also in Appendix, we describe numerical methods.
In Sec.~IV, we mention the extended Nagaoka ferromagnetism at zero temperature.
In Sec.~V, we survey ordering processes in the overall temperature range
by analyzing various quantities,
such as energy and specific heat.
In Sec.~VI, we focus on the magnetic property at low temperature.
In Sec.~VII, we investigate the dependence of spin correlation functions on temperature and distance.
Section~VIII is devoted to summary and discussion.

\section{Model for ferromagnetism}

\begin{figure}[t] 
\centering
\includegraphics[clip,scale=1.0]{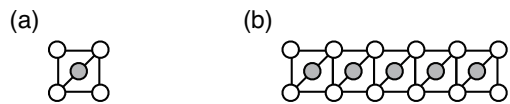}
\caption{
(a) Lattice unit structure to build an extended lattice.
A four-site plaquette (open circles) is called subsystem, which represent a main frame.
A center site (solid circle) is regarded as a particle bath.
Solid lines denote the hopping connection between two sites.
(b) Extended lattice built by arranging the units in one direction.
}
\label{Fig_lattice}
\end{figure}

We have studied a Hubbard model including sites that work as the particle bath
\cite{Miyashita2008,Onishi2014}.
Let us briefly explain how the ground state changes between
Mott AFM and Nagaoka FM states.
As a unit, we take a five-site system
depicted in Fig.~\ref{Fig_lattice}(a).
The system consists of two parts:
we call sites denoted by open circles ``subsystem'',
which represent a main frame,
and a shaded circle ``center site'',
which works as a particle bath.
We consider a concept of the Nagaoka ferromagnetism in the subsystem.
For this purpose,
we set the number of electrons to that of sites in the subsystem,
and control the distribution of electrons by the chemical potential at the center site.
For instance, we consider the case with four electrons in the five-site system
depicted in Fig.~\ref{Fig_lattice}(a).
The subsystem is half-filled if no electron is at the center site,
so that the system is in the Mott AFM state.
When the center site captures an electron, i.e., the subsystem has one hole,
the subsystem is in the Nagaoka FM state where the total spin is $3/2$.
A saturated FM state is realized
when the total spin of the system $S_{\mathrm{tot}}$ is given by
$S_{\mathrm{tot}}=3/2+1/2=2$.
We note that the total spin of the subsystem $S_{\mathrm{sub}}$
and that of the center site $S_{\mathrm{c}}$ are not considered separately,
but they are approximately given by $S_{\mathrm{sub}}=3/2$ and $S_{\mathrm{c}}=1/2$.

Extended lattices are built by arranging the units,
as shown in Fig.~\ref{Fig_lattice}(b).
The Hamiltonian is explicitly given by
\begin{align}
  H
  =&
  -t \sum_{\langle i,j \rangle,\sigma} (c_{i\sigma}^{\dagger}c_{j\sigma}+{\mathrm{h.c.}})
  +U \sum_{i} n_{i\uparrow}n_{i\downarrow}
\nonumber \\
  &
  +\mu \sum_{i \in {\mathrm{center}}} (n_{i\uparrow}+n_{i\downarrow}),
\label{Eq_H}
\end{align}
where $c_{i\sigma}$ and $c_{i\sigma}^{\dag}$ are annihilation and creation operators, respectively,
of an electron with spin $\sigma(=\uparrow,\downarrow)$ at the site $i$,
$n_{i\sigma}=c_{i\sigma}^{\dagger}c_{i\sigma}$ is an electron number operator,
$t$ is the electron hopping,
$U$ is the on-site Coulomb interaction at all the sites,
and $\mu$ is the on-site energy only at the center sites,
which serves as the chemical potential of the particle bath.
Throughout the paper we take $t=1$ as the energy unit.
We use both open and periodic boundary conditions (OBC and PBC, respectively).
As mentioned above,
the total number of sites $N$ is the sum of
the number of sites in the subsystem $N^{\mathrm{sub}}$ and
that in the center sites $N^{\mathrm{c}}$, i.e.,
$N=N^{\mathrm{sub}}+N^{\mathrm{c}}$.
The number of electrons $N_{\mathrm{e}}$ is set equal to $N^{\mathrm{sub}}$
to have the half-filled situation in the subsystem when $\mu$ is large.

In the previous paper~\cite{Onishi2014},
we have confirmed that this mechanism for the itinerant ferromagnetism is realized
in the extended lattice in one dimension (1D), depicted in Fig.~\ref{Fig_lattice}(b),
in the thermodynamic limit.
We call this type of FM state ``extended Nagaoka FM state''.
In Sec.~IV,
we give a brief explanation about the extended Nagaoka ferromagnetism at zero temperature
as a reference for the present paper.
After that,
we focus on the temperature dependence of the model
in Secs.~V$\sim$VII.

\section{Method}

To study the ground state we use the Lanczos method.
In the present model (\ref{Eq_H}),
the number of electrons of up spin $N_{\uparrow}$
and that of electrons of down spin $N_{\downarrow}$ conserve,
so that
the ground state is obtained as the lowest energy state
with specified $(N_{\uparrow},N_{\downarrow})$.
The number of basis states, i.e., the dimension of the Hamiltonian matrix,
is given by
$\mathcal{M}={}_{N}C_{N_{\uparrow}} \cdot {}_{N}C_{N_{\downarrow}}$,
which becomes huge as the system size increases.
Concretely,
$\mathcal{M}=3,136$
for the system with $N=8$ and $N_{\uparrow}=N_{\downarrow}=3$,
$\mathcal{M}=213,444$
for $N=11$ and $N_{\uparrow}=N_{\downarrow}=4$,
$\mathcal{M}=4,008,004$
for $N=14$ and $N_{\uparrow}=N_{\downarrow}=5$,
and
$\mathcal{M}=153,165,376$
for $N=17$ and $N_{\uparrow}=N_{\downarrow}=6$.

In the present work,
our main focus is on properties at finite temperatures.
For this purpose, we use an exact diagonalization (ED) method to obtain
all eigenvalues and eigenvectors of the Hamiltonian and calculate the thermal average.
This is a straightforward method
to calculate the thermal average of any physical quantities at any temperatures numerically exactly.
However, the computation is limited to small size systems
because the matrix dimension grows exponentially with the system size,
even when we block-diagonalize the Hamiltonian
in terms of $(N_{\uparrow},N_{\downarrow})$
to reduce the matrix dimension for efficient calculations.
We use the ED method to analyze systems with $N=8$ and $N_{\mathrm{e}}=6$,
and with $N=11$ and $N_{\mathrm{e}}=8$.

In order to investigate finite-temperature properties
for larger systems beyond those handled by the ED method,
we apply a random vector method~\cite{Hams2000,Jin2021,Sugiura2013}
which will be explained in Appendix.

\section{Extended Nagaoka Ferromagnetism at zero temperature}

\begin{figure}[t] 
\centering
\includegraphics[clip,scale=0.65]{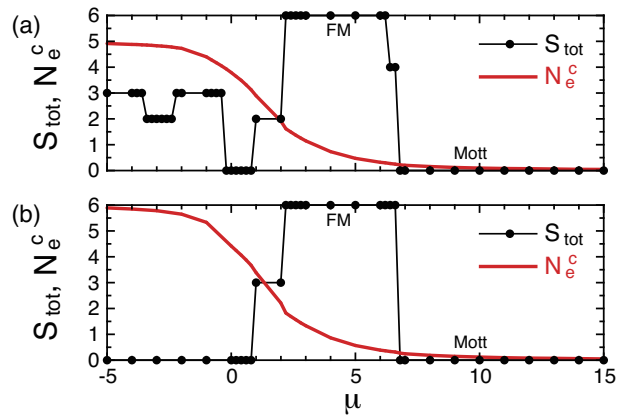}
\caption{
The total spin $S_{\mathrm{tot}}$ and
the number of electrons in the center sites $N_{\mathrm{e}}^{\mathrm{c}}$
as a function of $\mu$ at $U=500$ for
(a) the OBC lattice with $N=17$ and $N_{\mathrm{e}}=12$,
and for
(b) the PBC lattice with $N=18$ and $N_{\mathrm{e}}=12$.
Both lattices consist of 5 units.
}
\label{Fig_stot-nc_gs-1D}
\end{figure}

In the previous paper~\cite{Onishi2014}, we reported the realization of
the extended Nagaoka FM state at zero temperature
in lattices composed of units connected linearly [Fig.~\ref{Fig_lattice}(b)].
Analyzing the ground state with up to $N=110$ sites
by the Lanczos and density-matrix renormalization group methods,
we found that the extended Nagaoka FM state appears
in a certain range of the electron density in the subsystem.
We confirmed that the phase diagram little depends on the number of units,
which suggests that the mechanism holds in the thermodynamic limit.

In Fig.~\ref{Fig_stot-nc_gs-1D}(a),
we present typical results for $N=17$ and $N_{\mathrm{e}}=12$
as a reference for the present paper.
Here we plot the total spin $S_{\mathrm{tot}}$, evaluated by
\begin{equation}
  S_{\mathrm{tot}}(S_{\mathrm{tot}}+1) = \langle \bm{S}_{\mathrm{tot}}^{2} \rangle,
\label{Eq_Stot2}
\end{equation}
where $\bm{S}_{\mathrm{tot}}=\sum_{i}\bm{S}_{i}$
and $\langle\cdots\rangle$ denotes the expectation value in the ground state.
We numerically obtain the value of $\langle \bm{S}_{\mathrm{tot}}^{2} \rangle$
and estimate $S_{\mathrm{tot}}$ via
\begin{equation}
  S_{\mathrm{tot}} = (-1+\sqrt{1+4 \langle \bm{S}_{\mathrm{tot}}^{2} \rangle })/2.
\label{Eq_Stot}
\end{equation}
We also plot the number of electrons in the center sites,
\begin{equation}
  N_{\mathrm{e}}^{\mathrm{c}} = \sum_{i \in \mathrm{center},\sigma} \langle n_{i\sigma} \rangle.
\end{equation}
There appears a saturated FM state
when electrons are accommodated in the center sites,
i.e., holes are doped into the subsystem.
We stress again that we do not need to add or remove electrons one by one,
but we control the electron density by the chemical potential at the center site,
which remains well-defined in the thermodynamic limit.

We note that
an intermediate state of $S_{\mathrm{tot}}=N_{\mathrm{e}}/2-2$ appears in a narrow region
between the Mott state of $S_{\mathrm{tot}}=0$
and the extended Nagaoka FM state of $S_{\mathrm{tot}}=N_{\mathrm{e}}/2$,
as shown in Fig.~\ref{Fig_stot-nc_gs-1D}(a).
This state is found to be a bound state formed at corner sites
that are not connected to the center sites in the OBC.
This boundary effect is removed by taking the PBC,
where $S_{\mathrm{tot}}$ changes
from $0$ to $N_{\mathrm{e}}/2$ directly
without passing through intermediate $N_{\mathrm{e}}/2-2$,
as shown in Fig.~\ref{Fig_stot-nc_gs-1D}(b).

\begin{figure}[t] 
\centering
\includegraphics[clip,scale=0.65]{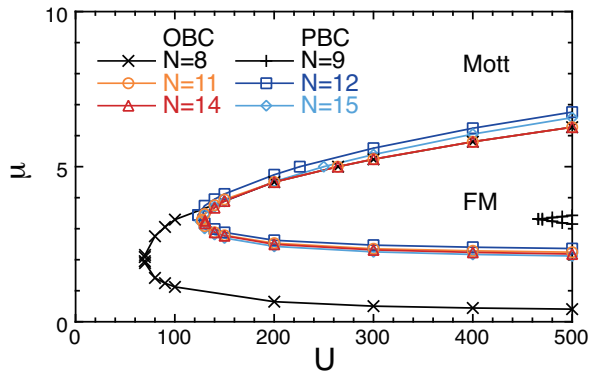}
\caption{
The ground-state phase diagram for 1D lattices.
The phase boundary to the saturated FM state is presented for various system sizes in the OBC and PBC.
}
\label{Fig_gspd-mu-u-1DOP}
\end{figure}

Figure~\ref{Fig_gspd-mu-u-1DOP} shows the ground-state phase diagram for the 1D lattice.
We plot the region where the saturated FM state of $S_{\mathrm{tot}}=N_{\mathrm{e}}/2$ appears
for various system sizes in the OBC and PBC.
To avoid making the phase diagram complicated,
we do not show phase boundaries for other values of $S_{\mathrm{tot}}$,
such as above mentioned $S_{\mathrm{tot}}=N_{\mathrm{e}}/2-2$ in the OBC.
We find that the phase boundaries of the systems of $N=8$ and 9 deviate from others,
while the system size dependence seems small for larger systems.
In this paper,
we concern the transition between Mott and FM states.
The phase boundary between them, i.e., the upper boundary,
does not depend much on the system size,
which supports that
the phase diagram includes 
the region of the saturated FM state in the thermodynamic limit,
as we proposed in the previous paper~\cite{Onishi2014}.

\section{Finite-temperature properties: Energy scales of charge and spin degrees of freedom}

To have an insight into finite-temperature properties of the extended Nagaoka ferromagnetism,
we investigate various physical quantities in 1D lattices,
such as energy, specific heat, and spin correlation function,
as a function of the temperature.
We mainly use the data for $N=8$ with the OBC,
obtained by the ED method,
and also study larger lattices by the random vector method.
We note that the overall trend does not change much with the system size,
as will be seen in Figs.~\ref{Fig_TCpeak_mu5} and \ref{Fig_C_n8-11-14_u500_mu5}.
This suggests that we can capture the general tendency by analyzing small systems of $N=8$.

\begin{figure}[t] 
\centering
\includegraphics[clip,scale=0.65]{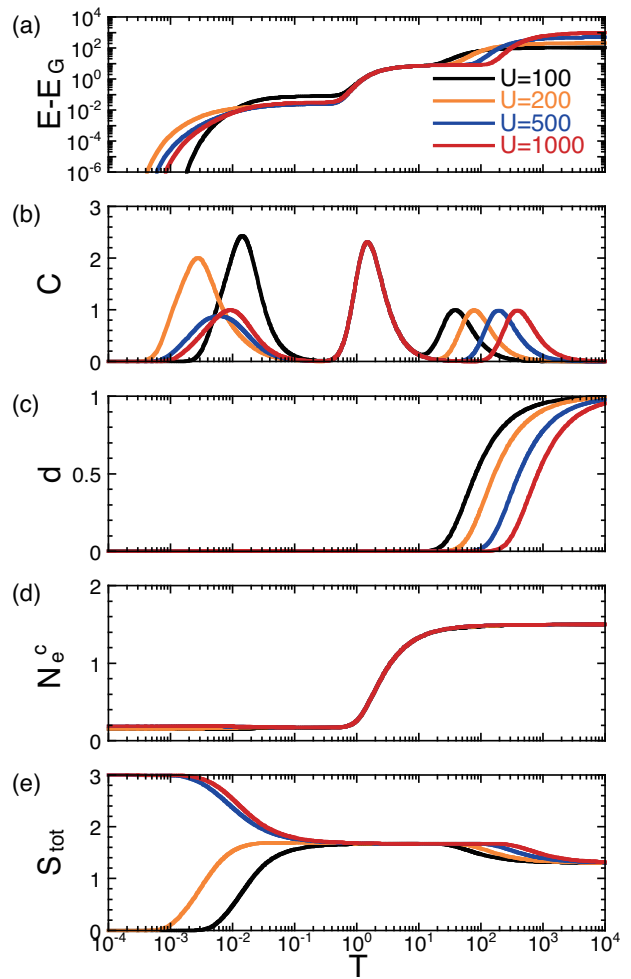}
\caption{
The temperature dependences of
(a) the energy, where the ground-state energy is subtracted,
(b) the specific heat,
(c) the double occupancy,
(d) the number of electrons in the center sites,
and
(e) the total spin
for typical $U$ at $\mu=5$.
$N=8$ and $N_{\mathrm{e}}=6$.
}
\label{Fig_E0_C_d_nc_stot_mu5}
\end{figure}

First, we study three steps of ordering processes
caused by $U$, $t$, and magnetic ordering,
producing characteristic peaks of the specific heat.
In Fig.~\ref{Fig_E0_C_d_nc_stot_mu5}(a),
we present the temperature dependence of the energy,
\begin{equation}
  E = \langle H \rangle_{T},
\end{equation}
for typical values of $U$ at $\mu=5$,
where $\langle \cdots \rangle_{T}$ denotes the thermal average at the temperature $T$.
We observe three marked changes in different temperature ranges
where the energy decreases significantly as the temperature goes down.
Note that the temperature axis is in the log scale,
and the different temperature ranges represent different scales of the temperature
by several orders of magnitude.
The energy axis is also in the log scale.
The characteristic temperatures where the energy decreases are clearly seen
as the peaks of the specific heat,
\begin{equation}
  C
  = \mathrm{d} E/ \mathrm{d} T
  = ( \langle H^{2} \rangle_{T} - \langle H \rangle_{T}^{2} )/T^{2},
\end{equation}
as shown in Fig.~\ref{Fig_E0_C_d_nc_stot_mu5}(b). 

In the high temperature limit which is higher than $U$,
electrons are randomly distributed in the system,
where electrons with different spins freely come to the same site,
while electrons with the same spin cannot occupy the site
due to the Pauli exclusion principle.
With lowering the temperature,
a change of the electron state is caused by the Coulomb interaction $U$
which gives the largest energy scale in the present model.
Because of $U$,
electrons with different spins avoid coming to the same site to lower the energy,
so that the energy drops at a temperature of the order of $U$.
Indeed, the characteristic temperature in this high temperature regime
increases with increasing $U$ monotonously.
This change of the electron state is actually confirmed by measuring the double occupancy,
\begin{equation}
  d = \sum_{i} \langle n_{i\uparrow}n_{i\downarrow} \rangle_{T},
\end{equation}
as shown in Fig.~\ref{Fig_E0_C_d_nc_stot_mu5}(c).
We find that the double occupancy changes at the temperature
corresponding to the peak of the specific heat seen in Fig.~\ref{Fig_E0_C_d_nc_stot_mu5}(b).

With further decreasing the temperature,
we observe a peak of the specific heat at an intermediate temperature,
as shown in Fig.~\ref{Fig_E0_C_d_nc_stot_mu5}(b).
The position of this second peak is independent of $U$,
located at around $T \simeq 1.5 \times 10^{0}$,
since curves of different $U$ coincide with each other.
This indicates that the relevant energy scale is of the order of $t$ and $\mu$.
With positive $\mu$,
electrons avoid occupying the center sites to decrease the on-site energy,
while the electron hopping brings an electron to the center sites to gain the kinetic energy.
By the balance of them,
an optimal spatial electron distribution is formed,
where the number of electrons in the center sites has an optimal value.
This optimization does not strongly depend on $U$,
as shown in Fig.~\ref{Fig_E0_C_d_nc_stot_mu5}(d).

Below this temperature,
charge degrees of freedom are effectively frozen.
However, there still exist spin degrees of freedom.
Its contribution to the energy is much smaller than
those of charge degrees of freedom.
Thus, changes of magnetic properties occur at low temperatures,
as is generally known.

Now we study the magnetic property of the present model,
although the low temperature causes difficulties in the numerical calculation.
To characterize the magnetic property,
we investigate the temperature dependence of the total spin $S_{\mathrm{tot}}$,
as shown in Fig.~\ref{Fig_E0_C_d_nc_stot_mu5}(e).
In the high temperature limit,
$S_{\mathrm{tot}}$ takes a constant value around 1.30 regardless of $U$.
This value is explained as follows.
In general,
\begin{equation}
  \langle \bm{S}_{\mathrm{tot}}^{2} \rangle_{T}
  =
  3 \langle (\sum_{i}S_{i}^{z})^{2} \rangle_{T},
\end{equation}
since the spin space is isotropic in the present model.
In the high temperature limit,
electrons freely move with keeping the Pauli exclusion principle,
so that the value in the high temperature limit
$\langle \bm{S}_{\mathrm{tot}}^{2} \rangle_{\infty}$
is evaluated by counting the number of combinations,
\begin{align}
  \langle \bm{S}_{\mathrm{tot}}^{2} \rangle_{\infty}
  =&
  3 \sum_{N_{\uparrow}=0}^{N_{\mathrm{e}}}
  N_{\uparrow}^{2} \cdot {}_{N}C_{N_{\uparrow}} \cdot {}_{N}C_{N_{\mathrm{e}}-N_{\uparrow}}
\nonumber \\
  &
  -3 N_{\mathrm{e}} \sum_{N_{\uparrow}=0}^{N_{\mathrm{e}}}
  N_{\uparrow} \cdot {}_{N}C_{N_{\uparrow}} \cdot {}_{N}C_{N_{\mathrm{e}}-N_{\uparrow}}
\nonumber \\
  &
  +\frac{3}{4}N_{\mathrm{e}}^{2},
\end{align}
where 
$S_{\mathrm{tot}}^{z}=(N_{\uparrow}-N_{\downarrow})/2$ and $N_{\mathrm{e}}=N_{\uparrow}+N_{\downarrow}$ are used.
The total spin in the high temperature limit $S_{\mathrm{tot}}^{\infty}$ is given by
substituting $\langle \bm{S}_{\mathrm{tot}}^{2} \rangle_{\infty}$ to Eq.~(\ref{Eq_Stot}).
For the case of $N=8$ and $N_{\mathrm{e}}=6$,
we obtain
$\langle \bm{S}_{\mathrm{tot}}^{2} \rangle_{\infty} = 3$ and $S_{\mathrm{tot}}^{\infty} = 1.30$,
which agrees with results in Fig.~\ref{Fig_E0_C_d_nc_stot_mu5}(e).

At the intermediate temperature of the order of $t$ and $\mu$,
$S_{\mathrm{tot}}$ takes a constant value around 1.68.
Considering that the double occupancy is prohibited and spins freely fluctuate,
the total spin in the intermediate temperature regime
$S_{\mathrm{tot}}^{\mathrm{IT}}$
is evaluated as
\begin{equation}
  S_{\mathrm{tot}}^{\mathrm{IT}} = (-1+\sqrt{1+3N_{\mathrm{e}}})/2,
\end{equation}
which is obtained by assuming
$\langle \bm{S}_{i} \cdot \bm{S}_{j} \rangle_{T}=0$ $(i \neq j)$
in Eq.~(\ref{Eq_Stot}).
For $N=8$ and $N_{\mathrm{e}}=6$,
we obtain
$S_{\mathrm{tot}}^{\mathrm{IT}}=1.68$,
which agrees with results in Fig.~\ref{Fig_E0_C_d_nc_stot_mu5}(e).

As the temperature decreases in the low temperature regime,
$S_{\mathrm{tot}}$ exhibits distinct changes depending on $U$,
and eventually converges to the value of the ground state.
We note that when the ground state has
$S_{\mathrm{tot}}=0$ for $U=100$ and 200,
the temperature exhibiting the change of $S_{\mathrm{tot}}$ decreases with $U$.
In contrast,
it increases with $U$
when the ground state has $S_{\mathrm{tot}}=N_{\mathrm{e}}/2=3$ for $U=500$ and 1000.
This difference depends on how the magnetic correlation develops,
which we will discuss in the next section.

\begin{figure}[t] 
\centering
\includegraphics[clip,scale=0.65]{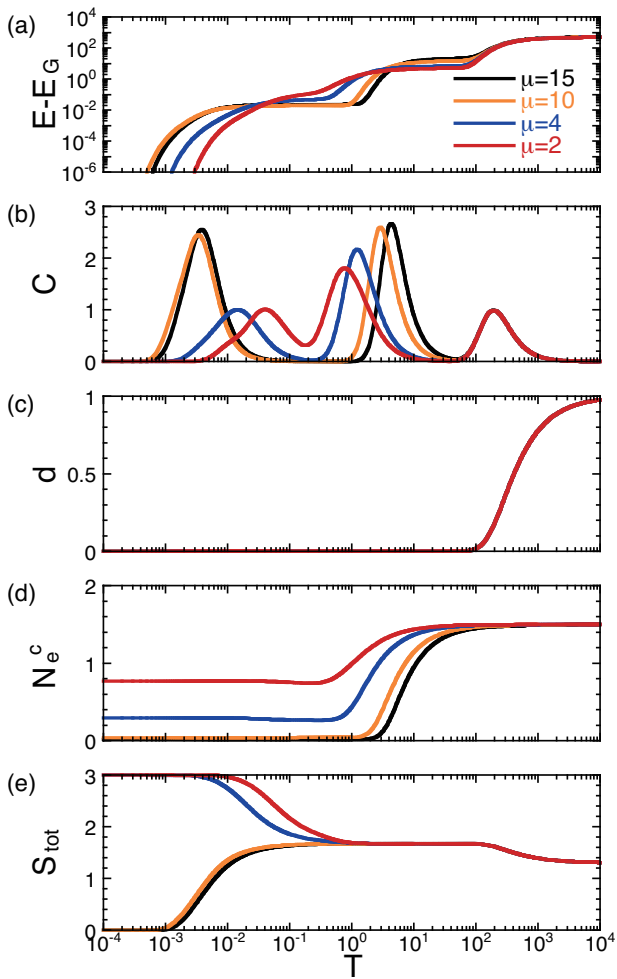}
\caption{
The temperature dependences of
(a) the energy, where the ground-state energy is subtracted,
(b) the specific heat,
(c) the double occupancy,
(d) the number of electrons in the center sites,
and
(e) the total spin
for typical $\mu$ at $U=500$.
$N=8$ and $N_{\mathrm{e}}=6$.
}
\label{Fig_E0_C_d_nc_stot_u500}
\end{figure}

Thus far we have focused on the $U$ dependence.
Here, we complementarily study the $\mu$ dependence
to obtain a deeper understanding of three steps of ordering processes
that occur in different temperature ranges.
In Figs.~\ref{Fig_E0_C_d_nc_stot_u500}(a) and \ref{Fig_E0_C_d_nc_stot_u500}(b),
we show the temperature dependences of the energy and the specific heat, respectively,
for typical $\mu$ at $U=500$.
We observe again that the energy decreases significantly in different temperature ranges,
causing the peaks of the specific heat.
The high temperature peak of the specific heat is located at around $T\simeq 1.9 \times 10^{2}$
independently of $\mu$.
The double occupancy drops there,
as shown in Fig.~\ref{Fig_E0_C_d_nc_stot_u500}(c).
This fact clearly indicates that the change of the electron state
in this high temperature regime is governed by $U$.

As seen in Fig.~\ref{Fig_E0_C_d_nc_stot_u500}(b),
the other two peaks of the specific heat
at intermediate and low temperatures move with $\mu$.
We find that the intermediate temperature peak shifts to the high temperature side monotonously
with increasing $\mu$.
This behavior is natural because the optimal spatial electron distribution is caused by $\mu$.
There, we observe that
the number of electrons in the center sites becomes large with decreasing $\mu$,
as shown in Fig.~\ref{Fig_E0_C_d_nc_stot_u500}(d),
as it should be.
On the other hand,
the low temperature peak is of magnetic origin.
In fact,
we find distinctive behavior of the total spin,
as shown in Fig.~\ref{Fig_E0_C_d_nc_stot_u500}(e).
With decreasing $\mu$,
the temperature exhibiting the change of $S_{\mathrm{tot}}$ decreases
when the ground state has $S_{\mathrm{tot}}=0$ for $\mu=15$ and 10,
while it turns to increase
when the ground state has $S_{\mathrm{tot}}=N_{\mathrm{e}}/2=3$ for $\mu=4$ and 2.
This opposite change across the quantum phase transition is similar
to what we observed with varying $U$ in Fig.~\ref{Fig_E0_C_d_nc_stot_mu5}.

\section{Magnetic phase diagram}

As observed in Fig.~\ref{Fig_E0_C_d_nc_stot_mu5}(e),
and also in the ground-state phase diagram in Fig.~\ref{Fig_gspd-mu-u-1DOP},
with varying $U$ at $\mu=5$,
the ground state is the Mott state of $S_{\mathrm{tot}}=0$ for small $U$
and the saturated FM state of $S_{\mathrm{tot}}=N_{\mathrm{e}}/2$ for large $U$,
and there is a quantum phase transition in between them.
The transition point is at $U_{\mathrm{c}}=264.3$
for $N=8$ and $N_{\mathrm{e}}=6$.
In the following, we investigate properties at low temperatures in three ranges of $U$ at $\mu=5$:
the Mott ground-state regime,
the FM ground-state regime,
and a region near the quantum phase transition.

\subsection{Mott ground-state regime}

\begin{figure}[t] 
\centering
\includegraphics[clip,scale=0.65]{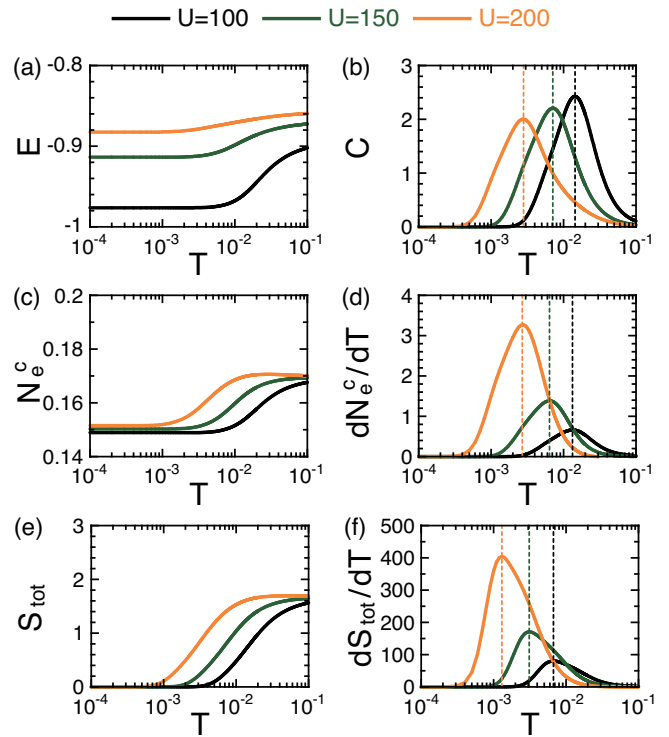}
\caption{
Low temperature properties for typical $U$ at $\mu=5$
where the ground state is the Mott state of $S_{\mathrm{tot}}=0$.
(a) The energy,
(b) the specific heat,
(c) the number of electrons in the center sites $N_{\mathrm{e}}^{\mathrm{c}}$,
(d) the temperature derivative of $N_{\mathrm{e}}^{\mathrm{c}}$,
(e) the total spin $S_{\mathrm{tot}}$,
and
(f) the temperature derivative of $S_{\mathrm{tot}}$
as a function of the temperature.
Vertical dotted lines in (b), (d), and (f) denote the peak positions.
$N=8$ and $N_{\mathrm{e}}=6$.
}
\label{Fig_E_nc_stot_DT_mu5_AFM}
\end{figure}

As well-known, assuming that each site is occupied by one electron,
the coupling between spins in neighboring sites is described by the AFM exchange interaction,
\begin{equation}
  J_{\mathrm{AFM}}(\bm{S}_{i}\cdot\bm{S}_{j}-1/4), \ \
  J_{\mathrm{AFM}} = \frac{4t^{2}}{U},
\label{Eq_JAFM}
\end{equation}
which is derived by the second-order perturbation with respect to the electron hopping $t$
in the strong-coupling limit.
Thus we expect that
AFM correlations grow at the temperature
corresponding to $J_{\mathrm{AFM}}$, being proportional to $1/U$.

In Figs.~\ref{Fig_E_nc_stot_DT_mu5_AFM}(a) and \ref{Fig_E_nc_stot_DT_mu5_AFM}(b),
we present the energy and the specific heat, respectively,
for typical values of $U$ at $\mu=5$ where the ground state is the Mott state of $S_{\mathrm{tot}}=0$.
Here we plot the energy itself without subtracting the ground-state energy in the linear scale.
According to the $U$ dependence of $J_{\mathrm{AFM}}$ in Eq.~(\ref{Eq_JAFM}),
as $U$ increases, the energy reduction of magnetic origin at low temperature becomes small,
and the peak of the specific heat shifts toward the low temperature side.
That is, the magnetic energy scale becomes small with increasing $U$ in the Mott ground-state regime.

Figures~\ref{Fig_E_nc_stot_DT_mu5_AFM}(c) and \ref{Fig_E_nc_stot_DT_mu5_AFM}(d)
show, respectively, the number of electrons in the center sites $N_{\mathrm{e}}^{\mathrm{c}}$
and its temperature derivative.
As we discussed in Sec.~V,
an optimal spatial electron distribution is mostly formed
at the intermediate temperature of the order of $10^{-1}$.
However, $N_{\mathrm{e}}^{\mathrm{c}}$ is further reduced at low temperature.
This indicates that charge degrees of freedom are not completely frozen yet,
but a fine adjustment occurs due to magnetic properties.
Note that the reduction of $N_{\mathrm{e}}^{\mathrm{c}}$ indicates that
the electron filling in the subsystem increases
to approach the half-filling situation of the subsystem,
which is favorable to gain the magnetic energy
through the AFM exchange interaction in the subsystem.
This is a kind of spin-charge coupling.
Comparing with Figs.~\ref{Fig_E_nc_stot_DT_mu5_AFM}(b) and \ref{Fig_E_nc_stot_DT_mu5_AFM}(d),
the peak positions of the specific heat and $\mathrm{d}N_{\mathrm{e}}^{\mathrm{c}}/\mathrm{d}T$
agree with each other,
as denoted by vertical dotted lines.

As shown in Fig.~\ref{Fig_E_nc_stot_DT_mu5_AFM}(e),
$S_{\mathrm{tot}}$ changes correspondingly with
$N_{\mathrm{e}}^{\mathrm{c}}$ in Fig.~\ref{Fig_E_nc_stot_DT_mu5_AFM}(c),
indicating a spin-charge coupling.
We find in Fig.~\ref{Fig_E_nc_stot_DT_mu5_AFM}(f) that
the peak of $\mathrm{d}S_{\mathrm{tot}}/\mathrm{d}T$ locates at
a slightly lower temperature comparing with those of
the specific heat and $\mathrm{d}N_{\mathrm{e}}^{\mathrm{c}}/\mathrm{d}T$.
This indicates that
the charge state changes at a higher temperature
to realize a fine-adjusted electron distribution,
and the spin state goes to the ground state at a lower temperature.
The spin contribution to the specific heat at the lower temperature
is not identified as a separate peak,
since it is close to the peak due to the charge contribution at the higher temperature.

\subsection{FM ground-state regime}

\begin{figure}[t] 
\centering
\includegraphics[clip,scale=0.65]{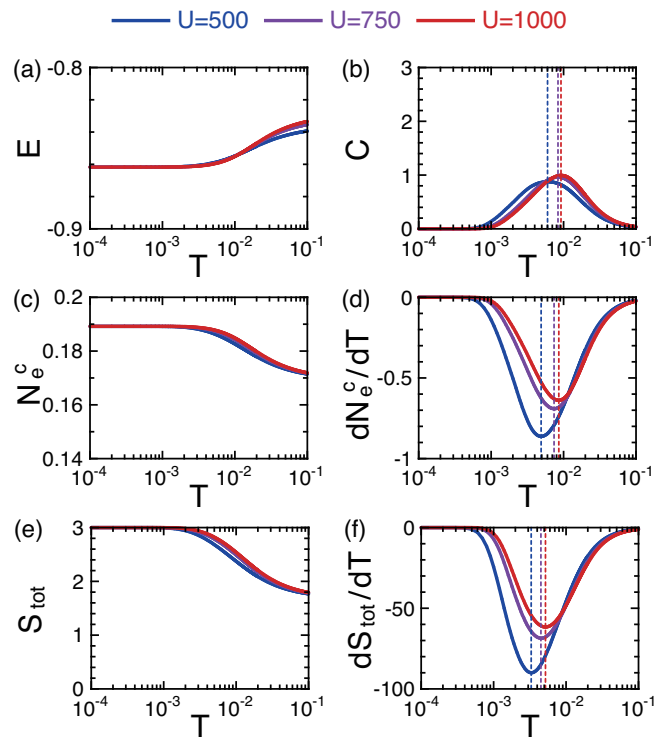}
\caption{
Low temperature properties for typical $U$ at $\mu=5$
where the ground state is the FM state of $S_{\mathrm{tot}}=N_{\mathrm{e}}/2$.
(a) The energy,
(b) the specific heat,
(c) the number of electrons in the center sites $N_{\mathrm{e}}^{\mathrm{c}}$,
(d) the temperature derivative of $N_{\mathrm{e}}^{\mathrm{c}}$,
(e) the total spin $S_{\mathrm{tot}}$,
and
(f) the temperature derivative of $S_{\mathrm{tot}}$
as a function of the temperature.
Vertical dotted lines in (b), (d), and (f) denote the peak positions.
$N=8$ and $N_{\mathrm{e}}=6$.
}
\label{Fig_E_nc_stot_DT_mu5_FM}
\end{figure}

We move on to the regime where the ground state is
the saturated FM state of $S_{\mathrm{tot}}=N_{\mathrm{e}}/2$.
In Fig.~\ref{Fig_E_nc_stot_DT_mu5_FM}(a),
we present the energy.
It appears that the ground-state energy is independent of $U$.
This is because electrons with parallel spin alignment do not occupy the same site
due to the Pauli exclusion principle,
so that the Coulomb interaction does not affect the ground-state energy.

Figure~\ref{Fig_E_nc_stot_DT_mu5_FM}(b) shows the specific heat.
We find that
the peak of the specific heat shifts to the high temperature side
with increasing $U$,
indicating that the FM state is stabilized by $U$.
The $U$ dependence of the peak shift seems rather gentle in comparison with that in the Mott ground-state regime
in Fig.~\ref{Fig_E_nc_stot_DT_mu5_AFM}(b).
The insensitivity to $U$ suggests that
the mechanism of the extended Nagaoka ferromagnetism is robustly realized
and its stability does not depend strongly on $U$.

As shown in Fig.~\ref{Fig_E_nc_stot_DT_mu5_FM}(c),
$N_{\mathrm{e}}^{\mathrm{c}}$ increases at low temperature.
Following the increase of $N_{\mathrm{e}}^{\mathrm{c}}$,
$S_{\mathrm{tot}}$ also increases,
as seen in Fig.~\ref{Fig_E_nc_stot_DT_mu5_FM}(e).
This is again a spin-charge coupling,
but the changes of $N_{\mathrm{e}}^{\mathrm{c}}$ and $S_{\mathrm{tot}}$
are opposite to those in the Mott ground-state regime
where $N_{\mathrm{e}}^{\mathrm{c}}$ and $S_{\mathrm{tot}}$ decrease with decreasing the temperature.
That is, more holes are introduced into the subsystem,
so that the hole motion in the subsystem is enhanced,
which works positively for the realization of the extended Nagaoka ferromagnetism.
The peak positions of $\mathrm{d}N_{\mathrm{e}}^{\mathrm{c}}/\mathrm{d}T$
are close to those of the specific heat
[Figs.~\ref{Fig_E_nc_stot_DT_mu5_FM}(b) and \ref{Fig_E_nc_stot_DT_mu5_FM}(d)],
and those of $\mathrm{d}S_{\mathrm{tot}}/\mathrm{d}T$
move to the low temperature side
[Figs.~\ref{Fig_E_nc_stot_DT_mu5_FM}(b) and \ref{Fig_E_nc_stot_DT_mu5_FM}(f)],
similarly to what we observed in Fig.~\ref{Fig_E_nc_stot_DT_mu5_AFM}.

\subsection{Quantum phase transition}

\begin{figure}[t] 
\centering
\includegraphics[clip,scale=0.65]{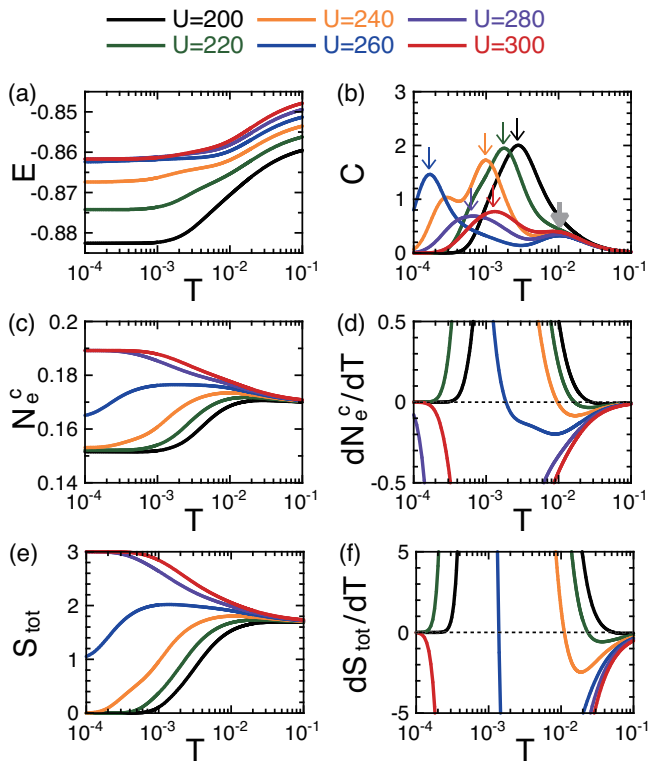}
\caption{
Low temperature properties for typical $U$ at $\mu=5$
near the transition between the Mott and FM ground states
at $U_{\mathrm{c}}=264.3$
for $N=8$ and $N_{\mathrm{e}}=6$.
(a) The energy,
(b) the specific heat,
(c) the number of electrons in the center sites $N_{\mathrm{e}}^{\mathrm{c}}$,
(d) the temperature derivative of $N_{\mathrm{e}}^{\mathrm{c}}$,
(e) the total spin $S_{\mathrm{tot}}$,
and
(f) the temperature derivative of $S_{\mathrm{tot}}$
as a function of the temperature.
In (b), positions of main peaks are denoted by thin arrows,
while a gray thick arrow is to mark an additional peak around $T\simeq 0.01$.
}
\label{Fig_E_nc_stot_DT_mu5_FM-AFM-boundary}
\end{figure}

Now we discuss finite-temperature properties near the quantum phase transition.
Figures~\ref{Fig_E_nc_stot_DT_mu5_FM-AFM-boundary}(a) and \ref{Fig_E_nc_stot_DT_mu5_FM-AFM-boundary}(b)
present the energy and the specific heat, respectively.
The peak of the specific heat
moves toward the low temperature side as $U$ increases below $U_{\mathrm{c}}$,
and it turns to move toward the high temperature side above $U_{\mathrm{c}}$,
as already seen in Figs.~\ref{Fig_E_nc_stot_DT_mu5_AFM}(b) and \ref{Fig_E_nc_stot_DT_mu5_FM}(b).
Moreover,
we find an additional peak at around $T \simeq 0.01$,
denoted by a gray thick arrow,
when $U$ is close to $U_{\mathrm{c}}$.
This peak should be attributed to the competition of different orders near the quantum phase transition.
We will discuss this competition by investigating the spin correlation
in Sec.~VII.

We find an additional peak in the low temperature side of the main peak for $U=240$,
which comes from discrete energy levels due to the finite-size effect,
the analysis of which strays from the main subject in the present paper
and we do not discuss them.

As seen in Figs.~\ref{Fig_E_nc_stot_DT_mu5_FM-AFM-boundary}(c) and \ref{Fig_E_nc_stot_DT_mu5_FM-AFM-boundary}(e),
respectively,
$N_{\mathrm{e}}^{\mathrm{c}}$ and $S_{\mathrm{tot}}$
simply increase with decreasing the temperature
for $U=300$ and 280,
where the system has the FM ground state.
Below $U_{\mathrm{c}}$,
for $U=240$ and 220,
where the system still remains near the transition point,
we find that
$N_{\mathrm{e}}^{\mathrm{c}}$ and $S_{\mathrm{tot}}$ first increase,
while they turn to decrease.
The increase of $S_{\mathrm{tot}}$ in the early stage,
i.e., the enhancement of the FM correlation,
is a reminiscent of the FM ground state above $U_{\mathrm{c}}$.
The AFM correlation turns to be large,
and eventually the system reaches the Mott ground state.
This type of inversion of AFM and FM correlations is a characteristic property of itinerant electrons.
In order to see this non-monotonic behavior,
we plot the derivatives in a magnified scale
in Figs.~\ref{Fig_E_nc_stot_DT_mu5_FM-AFM-boundary}(d) and \ref{Fig_E_nc_stot_DT_mu5_FM-AFM-boundary}(f).
Here we clearly see the sign change of the derivatives.

The ground state has $S_{\mathrm{tot}}=1$ for $U=260$,
as we see in Fig.~\ref{Fig_E_nc_stot_DT_mu5_FM-AFM-boundary}(e).
As mentioned in Sec.~III,
between the phases of $S_{\mathrm{tot}}=0$ and $S_{\mathrm{tot}}=N_{\mathrm{e}}/2$,
there is a short interval
of an intermediate state with $S_{\mathrm{tot}}=N_{\mathrm{e}}/2-2$ in the OBC.
We do not look into this fact in detail in the present analysis.
The presence of this intermediate state does not affect the global structure of a phase diagram discussed below.

\subsection{Phase diagram}

\begin{figure}[t] 
\centering
\includegraphics[clip,scale=0.65]{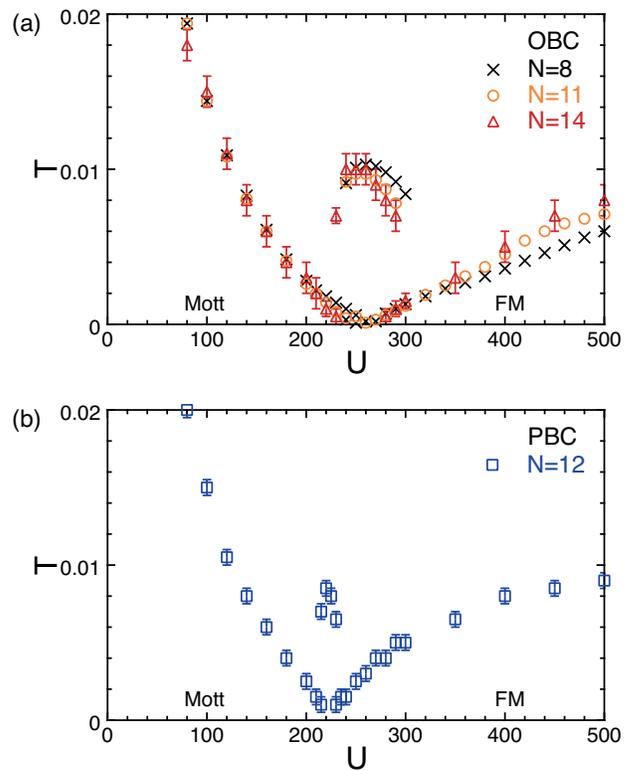}
\caption{
The peak temperatures of the specific heat as a function of $U$ at $\mu=5$
for 1D lattices in the (a) OBC and (b) PBC.
}
\label{Fig_TCpeak_mu5}
\end{figure}

In Fig.~\ref{Fig_TCpeak_mu5}, we present a kind of phase diagram,
where we plot the peak temperatures of the specific heat as a function of $U$ at $\mu=5$.
We show the results of the OBC and PBC separately
in Figs.~\ref{Fig_TCpeak_mu5}(a) and \ref{Fig_TCpeak_mu5}(b), respectively,
since there is a relatively large size dependence of the transition point in the PBC,
as shown in Fig.~\ref{Fig_gspd-mu-u-1DOP}.

Let us first focus on the results of the OBC in Fig.~\ref{Fig_TCpeak_mu5}(a).
Here we plot results of exact diagonalization for $N=8$ and 11,
together with those for $N=14$ obtained by the random vector method.
As $U$ approaches the quantum phase transition point $U_{\mathrm{c}} \simeq 264$,
the peak temperatures in both the Mott and FM ground-state regimes decrease,
and eventually merge at zero temperature,
leading to a V shape structure as usual quantum phase transitions.
In addition,
we find peaks at around $T \simeq 0.01$
in the vicinity of the quantum phase transition.
There develops a different kind of fluctuation
reflecting the competition of AFM and FM orderings.
The peaks around $T \simeq 0.01$ form a dome structure in the phase diagram.
We note that this dome structure
resembles the quasigap behavior of high-$T_{\mathrm{c}}$ superconductors.

As shown in Fig.~\ref{Fig_TCpeak_mu5}(b),
we also observe the V shape and dome structures in the PBC.
Thus, the microscopic origin of these peaks is not the boundary effect
but attributed to magnetic correlations.

Above we have focused on the phase diagram in the $(U,T)$ plane.
Instead, we may study the phase diagram in the $(\mu,T)$ plane.
In this case, we observe similar behavior such as V shape and dome structures
near the quantum phase transition between the Mott and FM ground states (not shown).

\section{Temperature denendence of spin correlation}

\begin{figure*}[t] 
\centering
\includegraphics[clip,scale=0.65]{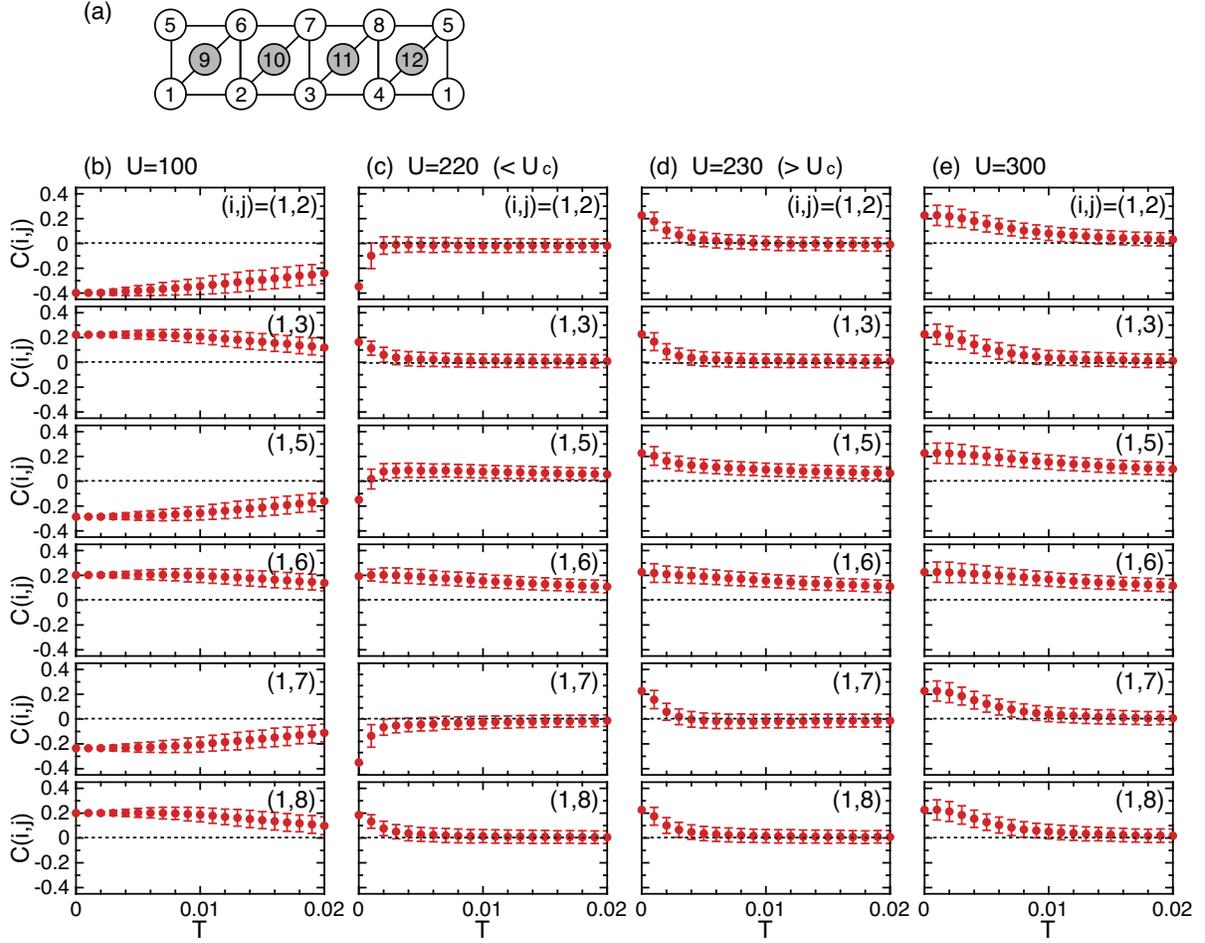}
\caption{
The spin correlation function $C(i,j)$ as a function of the temperature
for the 1D lattice in the PBC with 12 sites.
(a) The site numbering of the 12-site lattice.
$C(i,j)$ for
(b) $U=100$ well below $U_{\mathrm{c}} \simeq 225$,
(c) $U=220$ just below $U_{\mathrm{c}}$,
(d) $U=230$ just above $U_{\mathrm{c}}$,
and
(e) $U=300$ well above $U_{\mathrm{c}}$
at $\mu=5$.
The spin correlation is measured from the site $i=1$,
and the site pair $(i,j)$ is given in each panel.
The errorbars denote the standard deviation of the sampling data
in the random vector method.
The data of $T=0$ are obtained by the Lanczos method.
}
\label{Fig_cs_x3y1n12p_mu5}
\end{figure*}

\begin{figure*}[t] 
\centering
\includegraphics[clip,scale=0.65]{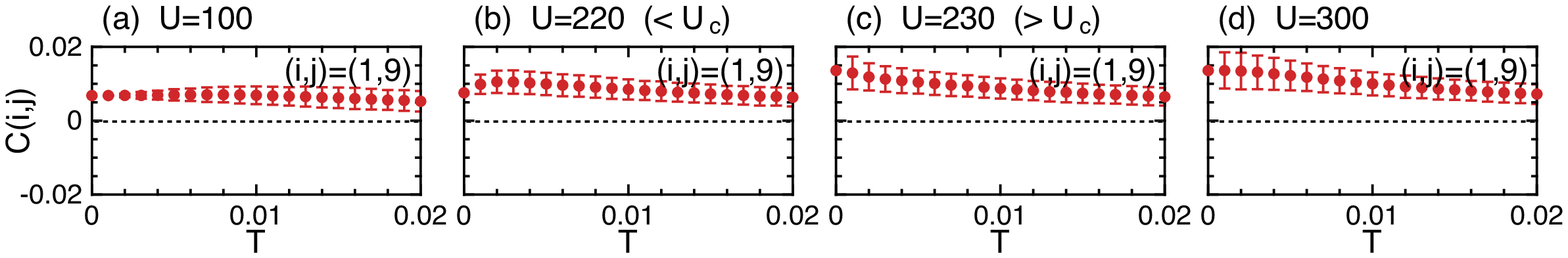}
\caption{
The spin correlation function $C(i,j)$ for the center site $j=9$ measured from the site $i=1$
for the 1D lattice in the PBC with 12 sites.
(a) $U=100$,
(b) $U=220$,
(c) $U=230$ ,
and
(d) $U=300$
at $\mu=5$,
of which the setting are the same as those of
Figs.~\ref{Fig_cs_x3y1n12p_mu5}(b),
\ref{Fig_cs_x3y1n12p_mu5}(c),
\ref{Fig_cs_x3y1n12p_mu5}(d),
and \ref{Fig_cs_x3y1n12p_mu5}(e),
respectively.
}
\label{Fig_cs-ctr_x3y1n12p_mu5}
\end{figure*}

As mentioned in the introductory part,
how magnetic order develops in itinerant ferromagnets as a function of the temperature is an interesting problem.
At $T=0$ all the spins are aligned parallel,
but at finite temperature the order is reduced.
To clarify the ordering process of itinerant electron spins from a microscopic viewpoint,
we measure the position dependence of the spin correlation function,
\begin{equation}
  C(i,j) = \langle \bm{S}_{i} \cdot \bm{S}_{j} \rangle_{T}.
\end{equation}
For the analysis of $C(i,j)$,
we use the PBC to avoid the boundary effect.
Here we take four values of $U$
near and away from the quantum phase transition point $U_{\mathrm{c}} \simeq 225$ at $\mu=5$
for a periodic lattice with 12 sites,
depicted in Fig.~\ref{Fig_cs_x3y1n12p_mu5}(a).
In the following we show $C(i,j)$ measured from the site $i=1$.

First, we focus on the spin correlation functions in the subsystem,
which are presented in Fig.~\ref{Fig_cs_x3y1n12p_mu5}.
If $U$ is well separated from $U_{\mathrm{c}} \simeq 225$,
the system has well-defined magnetic order.
For instance, at $U=100$ well below $U_{\mathrm{c}}$ [Fig.~\ref{Fig_cs_x3y1n12p_mu5}(b)],
the system has an AFM order,
which is short-ranged in the same way as the two-leg ladder AFM Heisenberg model of localized spins.
At $U=300$ well above $U_{\mathrm{c}}$ [Fig.~\ref{Fig_cs_x3y1n12p_mu5}(e)],
the system has a FM order.
In both cases, the spin orders at $T=0$ are reduced by the finite-temperature effect in a standard way,
where spin correlations at long distances decay quickly as the temperature increases.

Between them, a quantum phase transition occurs as found in the previous section,
around which the magnetic order appears at much low temperatures.
In Fig.~\ref{Fig_cs_x3y1n12p_mu5}(c),
we present $C(i,j)$ at $U=220$,
which is close to $U_{\mathrm{c}}$ and the ground state is the Mott state.
The system has an AFM order at very low temperature $T< 0.001$.
However, we find some peculiar behavior.
The spin correlation for $j=5$ shows non-monotonic temperature dependence and 
changes the sign at around $T \simeq 0.001$.
That is, the spin correlation is FM above $T > 0.001$,
changes its sign around $T\simeq 0.001$, and 
becomes AFM below $T<0.001$
which is consistent with the corresponding AFM Heisenberg model of localized spins.
Moreover,
although most of spin correlations are much reduced and quickly decay with the temperature
in comparison with those in Fig.~\ref{Fig_cs_x3y1n12p_mu5}(b),
the FM correlation of $j=6$ is robust and persists up to a relatively high temperature $T\simeq 0.05$,
as shown in Fig.~\ref{Fig_cs_x3y1n12p_mu5}(c) (shown up to $T=0.02$).

Figure~\ref{Fig_cs_x3y1n12p_mu5}(d) shows $C(i,j)$ at $U=230$
which is close to $U_{\mathrm{c}}$ and the ground state is the FM state.
The system has a FM order at very low temperature $T < 0.003$.
We observe similar behavior to the case of the Mott ground state at $U=220$ for $j=5$, and 6
at $T > 0.003$, i.e.,
FM correlations remain up to a relatively high temperature $T \simeq 0.05$.

The above observations suggest that
the formation of a short-range FM cluster (1-5-6-9) of itinerant electrons takes place.
That is, the FM spin alignment occurs via the hole motion in this cluster.
The growth of short-range FM order gives the peak of the specific heat,
causing the dome structure in Fig.~\ref{Fig_TCpeak_mu5}.
To clarify the spin state in this cluster,
we also measure the spin correlation function for $j=9$ which is a center site,
as shown in Fig.~\ref{Fig_cs-ctr_x3y1n12p_mu5}.
The spin correlation should be AFM for $U=100$ and 220
if we simply assume the AFM exchange interaction on the bond $(1,9)$,
but we find similar FM correlations
whether the ground state is FM or AFM.
This indicates that the FM correlation originates from the motion of itinerant electrons
in a short-range cluster (1-5-6-9),
which is a special property of the present itinerant FM system
in contrast to localized spin systems.

We consider that competition between this kind of growth of FM correlations and
AFM correlations due to the Mott mechanism near $U_{\mathrm{c}}$
causes the non-monotonic dependence of the spin correlation for $j=5$ and also
the peak of the specific heat at around $T \simeq 0.01$,
leading to the dome structure in Fig.~\ref{Fig_TCpeak_mu5}.

\begin{figure}[t] 
\centering
\includegraphics[clip,scale=0.65]{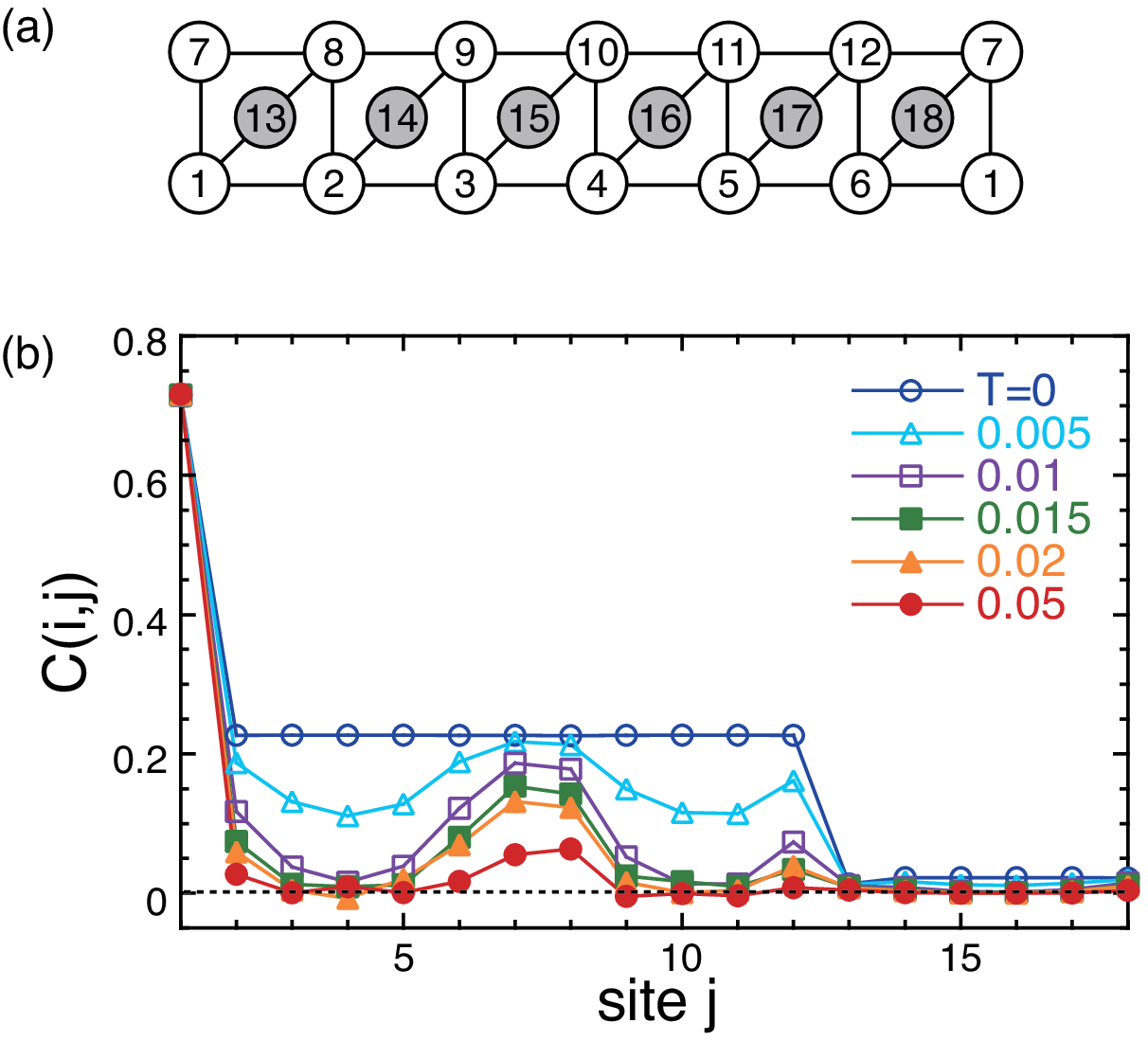}
\caption{
The spin correlation function $C(i,j)$ as a function of the site position
for the 1D lattice in the PBC with 18 sites.
(a) The site numbering of the 18-site lattice.
(b) $C(i,j)$ at several temperatures at $U=500$ and $\mu=5$.
The spin correlation is measured from the site $i=1$.
The data of $T=0$ are obtained by the Lanczos method.
}
\label{Fig_cs_x5y1n18p_u500_mu5}
\end{figure}

We further investigate a larger system with 18 sites
to clarify the position dependence of the spin correlation at long distances.
In Fig.~\ref{Fig_cs_x5y1n18p_u500_mu5},
we show $C(i,j)$ at $U=500$ and $\mu=5$,
where the ground state is the FM state of $S_{\mathrm{tot}}=N_{\mathrm{e}}/2$.
As the temperature increases,
the spin correlations at long distances decay fast.
On the other hand,
FM correlations in a short-range cluster persist up to relatively high temperature $T\simeq 0.05$.
Thus the formation of the FM cluster is not due to the finite-size effect
but inherent in the system of itinerant electrons.

\section{Summary and discussion}

In this paper, we studied finite-temperature properties of the model
for the extended Nagaoka ferromagnetism by using numerical methods.
In contrast to the original Nagaoka model,
we can deal with finite electron density by making use of the sites of the particle bath,
and thus, the thermodynamic limit is well-defined.

The mechanism of the alignment of spins in the present itinerant electron model is very different from
that in the localized spin Heisenberg model with local FM exchange interactions.
Since the present model exhibits FM and AFM ground states depending on the parameter $U$,
the temperature dependence of magnetic ordering is a matter of interest.
We surveyed ordering processes in the full range of the temperature.

The present model has three parameters $U$, $t$ and $\mu$.
Here we took $U \gg t$ and $t$ and $\mu$ in the same order,
since the FM state appears for sufficiently large $U$.
We found prominent peaks of the specific heat in three temperature regions
due to changes between characteristic electron states
(Figs.~\ref{Fig_E0_C_d_nc_stot_mu5} and \ref{Fig_E0_C_d_nc_stot_u500}).
The changes of the electron state are clearly identified by corresponding quantities,
such as the double occupancy, the electron density at the center site, and the total spin.
At high temperature $T \gg U$,
electrons distribute randomly
(complete random state),
where the restriction on the electron state is only the Pauli exclusion principle,
i.e., the double occupancy at the same site with the same spin.
At around $T \sim U$,
the double occupancy at the same site with the opposite spins becomes suppressed
(random state without double occupancy).
When $T$ decreases to the order of $t$ and $\mu$,
the electron distribution is affected by the lattice form,
i.e., the electron density at the particle bath in which $\mu$ is large is suppressed
and the subsystem becomes nearly half-filled
(itinerant paramagnetic state).
At much low temperature,
depending on the value of $U$ for a given $\mu$,
the system shows the FM state $(U>U_{\mathrm{c}})$ and the AFM state $(U<U_{\mathrm{c}})$,
(extended Nagaoka FM state and Mott AFM state, respectively).

Regarding magnetic properties,
we studied the details of the temperature dependence at very low temperature
where magnetic ordering occurs
(Figs.~\ref{Fig_E_nc_stot_DT_mu5_AFM}, \ref{Fig_E_nc_stot_DT_mu5_FM}, and \ref{Fig_E_nc_stot_DT_mu5_FM-AFM-boundary}).
In the present setup,
$U$ is the order of a few hundred
and the magnetic ordering takes place at around $T \sim 10^{-2}$, 
i.e., $T/U \sim 10^{-5}$.
This large difference between charge and magnetic orderings is a general feature
in the Hubbard model treatment,
and it causes the difficulty in multi-scale numerical calculations.
We used the random vector method (Appendix)
to handle with large lattices that are not available by the ED method.
If the dimension of Hilbert space is large enough,
single sample can produce the thermal average.
Indeed,
to study electron properties around $T > 10^{-2}$, 
single sample is enough to obtain the quantities precisely.
We confirmed this fact by checking the case with 5 samples.
However,
since the temperature is so low in the present case,
we needed sample average with a large number of samples,
e.g., 1000 samples.

The peak temperatures of the specific heat were given in Fig.~\ref{Fig_TCpeak_mu5}.
We found a V-shape structure as typically seen in the quantum phase transition.
Moreover, we found a dome structure around the critical $U$,
which resembles the quasi-gap behavior in the high-$T_{\mathrm{c}}$ superconductors.

To characterize magnetic orderings from a microscopic viewpoint,
we studied how the spin correlation develops with temperature
(Figs.~\ref{Fig_cs_x3y1n12p_mu5}, \ref{Fig_cs-ctr_x3y1n12p_mu5}, and \ref{Fig_cs_x5y1n18p_u500_mu5}).
We found that some local FM correlations are robust.
Even when the system is in the Mott AFM ground state regime below $U_{\mathrm{c}}$,
some neighboring spins that have AFM correlations in the ground state
show FM correlations in a certain temperature range,
which should be originated in the electron motion in a cluster.
This kind of competition occurs around the temperatures of the dome structure,
so that we attribute the peaks to this competition.

To study finite-temperature properties of the Hubbard model requires large computational resources.
Thus, in the present study, we only grasped the characteristics in small systems
where the finite-size scaling of quantities could not be examined sufficiently.
We expect that some more sophisticated methods, e.g., 
DMRG, tensor network method, etc.,
would extend the study in future.
It is also interesting future problems
to study finite-temperature properties of other models for itinerant ferromagnetism,
such as a flat-band model and a Kondo-lattice model with double exchange mechanism.

In the present study,
we considered the lattice build by one-dimensional arrangements of the unit structures.
However, it is preliminarily found that
the same kind of behavior has been found in two and three-dimensional arrangements
and the results of the present study are expected to takes place regardless of dimensions.

Finally we refer to related experimental systems for the realization of the extended Nagaoka FM state.
Recently, thanks to the development of experimental techniques
using cold atoms in optical lattices~\cite{Stecher2010,Okumura2011,Bloch2012},
and also,
the chemical synthesis of molecular magnets~\cite{Verdaguer1999,Coronado2013,Coronado2019}
and the fabrication of quantum dots~\cite{Dehollain2020},
quantum simulations of quantum lattice models have been made with high controllability.
In this context,
the Hubbard model is the simplest model of interacting fermions
because it consists of only transfer and on-site repulsion terms.
Thus, it is a good chance to realize the itinerant ferromagnetism in such systems.
If we prepare structures consisting of the main frame and the particle bath,
by controlling model parameters,
we can switch between FM and AFM states
and observe peculiar ordering processes in the vicinity of the quantum phase transition.

\begin{acknowledgments}
Computations were performed on supercomputers
at the Japan Atomic Energy Agency
and at the Institute for Solid State Physics, the University of Tokyo.
This work was in part supported by JSPS KAKENHI Grant Nos.~19K03678 and 18K03444,
and also by
the Elements Strategy Initiative Center for Magnetic Materials (ESICMM)
Grant No.~12016013 funded by the Ministry of Education,
Culture, Sports, Science and Technology (MEXT) of Japan.
\end{acknowledgments}

\section*{Appendix}

\begin{figure*}[t] 
\centering
\includegraphics[clip,scale=0.65]{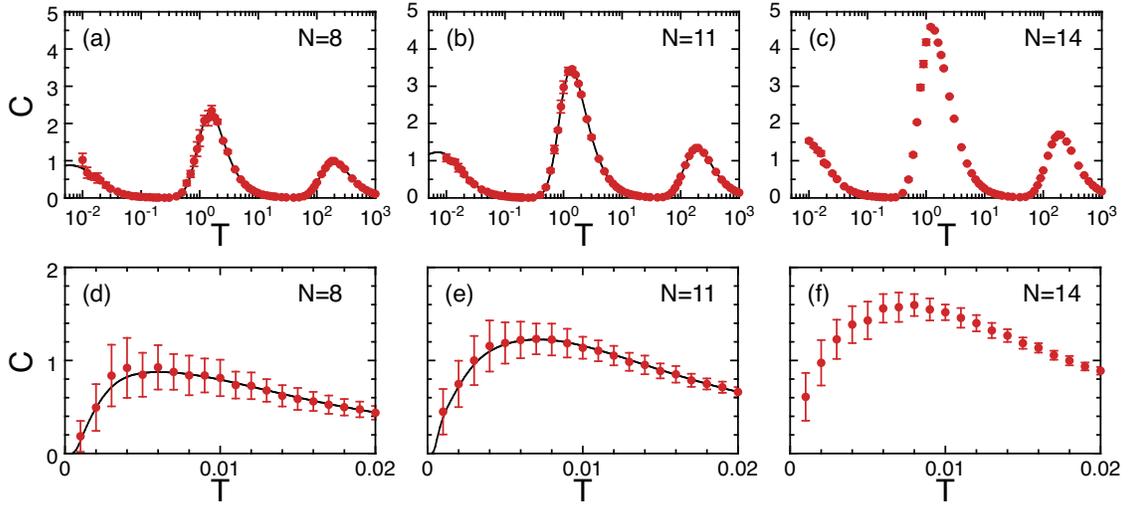}
\caption{
The specific heat as a function of the temperature for the 1D lattice in the OBC
at $U=500$ and $\mu=5$, obtained by the random vector method.
Top:
(a) $N=8$,
(b) $N=11$,
and
(c) $N=14$
in a wide range of the temperature,
obtained with 5 samples.
Bottom:
(d) $N=8$,
(e) $N=11$,
and
(f) $N=14$
at low temperatures where we study magnetic properties,
obtained with 100 samples.
The errorbars denote the standard deviation of the sampling data.
The solid curves denote the exact results by the ED method.
}
\label{Fig_C_n8-11-14_u500_mu5}
\end{figure*}

In this Appendix,
we briefly explain the random vector method~\cite{Hams2000,Jin2021,Sugiura2013},
and present some of numerical results to illustrate how we obtain results shown in the main text.

We first prepare an initial random vector $\vert \Phi \rangle$
of which each element is given by a gaussian distribution.
Then, we calculate a wavefunction for finite temperature $T$
by the imaginary-time evolution of the wavefunction,
\begin{equation}
  \vert \Phi_\beta \rangle = \mathrm{e}^{-\beta H/2}\vert \Phi \rangle,
\end{equation}
where $\beta=1/k_{\mathrm{B}}T$ is the inverse temperature
and $k_{\mathrm{B}}$ is the Boltzmann constant.
We use the Chebyshev polynomial representation of $\mathrm{e}^{-\beta H/2}$
to calculate $\mathrm{e}^{-\beta H/2}\vert \Phi \rangle$~\cite{Jin2021}.
The thermal average of a physical quantity $A$ is approximately obtained as
\begin{equation}
  \langle A \rangle_{T}
  =
  \frac{\mathrm{Tr}A\mathrm{e}^{-\beta H}}{\mathrm{Tr}\mathrm{e}^{-\beta H}}
  \simeq
  \frac{\langle \Phi_\beta \vert A \vert \Phi_\beta \rangle}{\langle \Phi_\beta \vert \Phi_\beta \rangle}.
\label{Eq_AT-approximate}
\end{equation}
Basically the random vector method is efficient at high temperatures,
since a random vector equally includes all eigenvectors
and it represents a thermal equilibrium state in the high temperature limit.
Regarding the accuracy at finite temperatures,
the difference between the last two terms in Eq.~(\ref{Eq_AT-approximate})
is proven to become exponentially small
with increasing the dimension of the Hamiltonian matrix.
However, at low temperatures, the difference becomes large and thus
we need to take the ensemble average of the last term
by using different initial random vectors $\vert \Phi \rangle$ to reduce the difference.

As mentioned above,
this method is efficient down to a certain temperature
even if we have only a few samples (even single sample)
to give a good estimation of physical quantities.
In Figs.~\ref{Fig_C_n8-11-14_u500_mu5}(a)-(c),
we show the specific heat in a wide range of the temperature at $U=500$ and $\mu=5$,
obtained with 5 samples.
Data of 5 samples have a very small distribution among them,
and they agree well with the exact ones obtained by the ED method for $N=8$ and 11
down to $T \sim 0.1$ (in units of $t$).
It is noted that the largest energy scale is given by $U$,
which is 500 in the present calculations.
We find that the method is efficient
even when the temperature is rather small
comparing with the largest energy scale $U$,
i.e., $T/U \sim 10^{-4}$.

However,
at much low temperatures below $T \sim 0.01$,
i.e., $T/U \lesssim 10^{-5}$,
where we study magnetic properties,
the self-averaging property due to the large Hilbert space is not enough to shape data.
In fact, the result of $N=8$ at $T=0.01$ with 5 samples does not agree with the exact one,
as seen in Fig.~\ref{Fig_C_n8-11-14_u500_mu5}(a).
Thus we need sample average by using a number of different initial random vectors.
In Figs.~\ref{Fig_C_n8-11-14_u500_mu5}(d)-(f),
we present the specific heat at low temperatures at $U=500$ and $\mu=5$,
obtained with 100 samples.
Although results of $N=8$ deviate from the ED results to some extent,
they are consistent within the errorbars.
The agreement with the ED results becomes better for $N=11$.
We also point out that
the distribution of the data of samples becomes small as the system size increases.
These tendencies come from the fact that the difference
between the last two terms in Eq.~(\ref{Eq_AT-approximate}) becomes small
as the size of the Hilbert space increases.
Therefore, we need fewer samples for larger systems
to obtain converged distributions to estimate the average and errorbar of quantities.



\begin{thebibliography}{99}

\bibitem{Heisenberg1928}
W. Heisenberg,
Z. Phys. \textbf{49}, 619 (1928).

\bibitem{Slater1936}
J. C. Slater,
Phys. Rev. \textbf{49}, 537 (1936).

\bibitem{Stoner1938}
E. C. Stoner,
Proc. R. Soc. Lond. A \textbf{165}, 372 (1938)

\bibitem{Slater1953}
J. C. Slater, H. Statz, and G. F. Koster,
Phys. Rev. \textbf{91}, 1323 (1953).

\bibitem{Goodenough1955}
J. B. Goodenough,
Phys. Rev. \textbf{100}, 564 (1955).

\bibitem{Kanamori1957a}
J. Kanamori,
Prog. Theor. Phys. \textbf{17}, 177 (1957).

\bibitem{Kanamori1957b}
J. Kanamori,
Prog. Theor. Phys. \textbf{17}, 197 (1957).

\bibitem{Goodenough1958}
J. B. Goodenough,
J. Phys. Chem. Solids \textbf{6}, 287 (1958).

\bibitem{Kanamori1959}
J. Kanamori,
J. Phys. Chem. Solids \textbf{10}, 87 (1959).

\bibitem{Kohn1999}
W.Kohn,
Rev. Mod. Phys. \textbf{71}, 1253 (1999).

\bibitem{Jones2015}
R. O. Jones,
Rev. Mod. Phys. \textbf{87}, 897 (2015).

\bibitem{Janak1979}
J. F. Janak,
Phys. Rev. B \textbf{20}, 2206 (1979).

\bibitem{Akai1990}
H. Akai, M. Akai, S. Bl\"ugel, B. Drittler, H. Ebert, K. Terakura, R. Zeller, and P. H. Dederichs,
Prog. Theor. Phys. Suppl. \textbf{101}, 11 (1990).

\bibitem{Moriya1973a}
T. Moriya and A. Kawabata,
J. Phys. Soc. Jpn. \textbf{34}, 639 (1973).

\bibitem{Moriya1973b}
T. Moriya and A. Kawabata,
J. Phys. Soc. Jpn. \textbf{35}, 669 (1973).

\bibitem{Moriya-book1985}
T. Moriya,
\textit{Spin Fluctuations in Itinerant Electron Magnetism}
(Springer, 1985).

\bibitem{Hubbard1963}
J. Hubbard,
Proc. R. Soc. Lond. A \textbf{276}, 238 (1963).

\bibitem{Kanamori1963}
J. Kanamori,
Prog. Theor. Phys. \textbf{30}, 275 (1963).

\bibitem{Gutzwiller1963}
M. C. Gutzwiller,
Phys. Rev. Lett. \textbf{10}, 159 (1963).

\bibitem{Nagaoka1966}
Y. Nagaoka,
Phys. Rev. \textbf{147}, 392 (1966).

\bibitem{Thouless1965}
D. J. Thouless,
Proc. Phys. Soc. \textbf{86}, 893 (1965).

\bibitem{Tasaki1989}
H. Tasaki,
Phys. Rev. B \textbf{40}, 9192 (1989).

\bibitem{Mielke1991a}
A. Mielke,
J. Phys. A: Math. Gen.\textbf{24}, L73 (1991).

\bibitem{Mielke1991b}
A. Mielke,
J. Phys. A: Math. Gen.\textbf{24}, 3311 (1991).

\bibitem{Mielke1992}
A. Mielke,
J. Phys. A: Math. Gen.\textbf{25}, 4335 (1991).

\bibitem{Tasaki1992}
H. Tasaki,
Phys. Rev. Lett. \textbf{69}, 1608 (1992).

\bibitem{Mielke1993}
A. Mielke and H. Tasaki,
Commun. Math. Phys. \textbf{158}, 341 (1993).

\bibitem{Tasaki1994}
H. Tasaki,
Phys. Rev. Lett. \textbf{73}, 1158 (1994).

\bibitem{Tasaki1996}
H. Tasaki,
J. Stat. Phys. \textbf{84}, 535 (1996).

\bibitem{Kubo1982}
K. Kubo,
J. Phys. Soc. Jpn. \textbf{51}, 782 (1982).

\bibitem{Kusakabe1994}
K. Kusakabe and H. Aoki,
Physica B \textbf{194-196}, 217 (1994).

\bibitem{Onishi2004}
H. Onishi and T. Hotta,
New J. Phys. \textbf{6}, 193 (2004).

\bibitem{Onishi2007a}
H. Onishi,
J. Magn. Magn. Mater. \textbf{310}, 790 (2007).

\bibitem{Onishi2007b}
H. Onishi,
Phys. Rev. B \textbf{76}, 014441 (2007).

\bibitem{Zener1951}
C. Zener,
Phys. Rev. \textbf{81}, 440 (1951).

\bibitem{Anderson1955}
P. W. Anderson and H. Hasegawa,
Phys. Rev. \textbf{100}, 675 (1955).

\bibitem{deGennes1960}
P. G. de Gennes,
Phys. Rev. \textbf{118}, 141 (1960).

\bibitem{Sigrist1991}
M. Sigrist, H. Tsunetsugu, and K. Ueda,
Phys. Rev. Lett. \textbf{67}, 2211 (1991).

\bibitem{Sigrist1992}
M. Sigrist, H. Tsunetsugu, K. Ueda, and T. M. Rice,
Phys. Rev. B \textbf{46}, 13838 (1992).

\bibitem{Tsunetsugu1997}
H. Tsunetsugu, M. Sigrist, and K. Ueda,
Rev. Mod. Phys. \textbf{69}, 809 (1997).

\bibitem{Nolting2003}
W. Nolting, W. Muller, and C. Santos,
J. Phys. A: Math. Gen. \textbf{36}, 9275 (2003).

\bibitem{Yamamoto2010}
S. J. Yamamoto and Q. Si,
PNAS \textbf{107}, 15704 (2010).

\bibitem{Takahashi1982}
M. Takahashi,
J. Phys. Soc. Jpn. \textbf{51}, 3475 (1982).

\bibitem{Doucot1989}
B. Doucot and X. G. Wen,
Phys. Rev. B \textbf{40}, 2719 (1989).

\bibitem{Riera1989}
J. A. Riera and A. P. Young,
Phys. Rev. B \textbf{40}, 5285 (1989).

\bibitem{Fang1989}
Y. Fang, A. E. Ruckenstein, E. Dagotto, and S. Schmitt-Rink,
Phys. Rev. B \textbf{40}, 7406 (1989).

\bibitem{Shastry1990}
B. S. Shastry, H. R. Krishnamurthy, and P. W. Anderson,
Phys. Rev. B \textbf{41}, 2375 (1990).

\bibitem{Suto1991}
A. S\"{u}t\H{o},
Commun. Math. Phys. \textbf{140}, 43 (1991).

\bibitem{Toth1991}
B. T\'oth,
Lett. Math. Phys. \textbf{22}, 321 (1991).

\bibitem{Hanisch1993}
T. Hanisch and E. M\"uller-Hartmann,
Ann. Phys. \textbf{2}, 381 (1993).

\bibitem{Hanisch1995}
T. Hanisch, B. Kleine, A. Ritzl, and E. M\"uller-Hartmann,
Ann. Phys. \textbf{4}, 303 (1995).

\bibitem{Sakamoto1996}
H. Sakamoto and K. Kubo,
J. Phys. Soc. Jpn. \textbf{65}, 3732 (1996).

\bibitem{Arita1998}
R. Arita, K. Kusakabe, K. Kuroki, and H. Aoki,
Phys. Rev. B \textbf{58}, R11833 (1998).

\bibitem{Daul1997}
S. Daul and R. M. Noack,
Z. Phys. B \textbf{103}, 293 (1997).

\bibitem{Daul1998}
S. Daul and R. M. Noack,
Phys. Rev. B \textbf{58}, 2635 (1998).

\bibitem{Kohno1997}
M. Kohno,
Phys. Rev. B \textbf{56}, 15015 (1997).

\bibitem{Watanabe1997a}
Y. Watanabe and S. Miyashita,
J. Phys. Soc. Jpn. \textbf{66}, 2123 (1997).

\bibitem{Watanabe1997b}
Y. Watanabe and S. Miyashita,
J. Phys. Soc. Jpn. \textbf{66}, 3981 (1997).

\bibitem{Watanabe1999}
Y. Watanabe and S. Miyashita,
J. Phys. Soc. Jpn. \textbf{68}, 3086 (1999).

\bibitem{Miyashita2008}
S. Miyashita,
Prog. Theor. Phys. \textbf{120}, 785 (2008).

\bibitem{Onishi2014}
H. Onishi and S. Miyashita,
Phys. Rev. B \textbf{90}, 224426 (2014).

\bibitem{Hams2000}
A. Hams and H. De Raedt,
Phys. Rev. E \textbf{62}, 4365 (2000).

\bibitem{Jin2021}
F. Jin, D. Willsch, M. Willsch, H. Lagemann, K. Michielsen, and H. De Raedt,
J. Phys. Soc. Jpn. \textbf{90}, 012001 (2021).

\bibitem{Sugiura2013}
S. Sugiura and A. Shimizu,
Phys. Rev. Lett. \textbf{111}, 010401 (2013).

\bibitem{Stecher2010}
J. von Stecher, E. Demler, M. D. Lukin, and A. M. Rey,
New J. Phys. \textbf{12}, 055009 (2010).

\bibitem{Okumura2011}
M. Okumura, S. Yamada, M. Machida, and H. Aoki,
Phys. Rev. A \textbf{83}, 031606(R) (2011).

\bibitem{Bloch2012}
I. Bloch, J. Dalibard, and S. Nascimb\`ene,
Nat. Phys. \textbf{8}, 267 (2012).

\bibitem{Verdaguer1999}
M. Verdaguer, A. Bleuzen, V. Marvaud, J. Vaissermann, M. Seuleiman, C. Desplanches, A. Suiller, C. Train, R. Garde, G. Gelly, C. Lomenech, I. Rosenman, P. Veillet, C. Cartier, and F. Villain,
Coord. Chem. Rev. \textbf{190-192}, 1023 (1999).

\bibitem{Coronado2013}
E. Coronado and G. Minguez Espallargas,
Chem. Soc. Rev. \textbf{42}, 1525 (2013).

\bibitem{Coronado2019}
E. Coronado,
Nat. Rev. Mat. \textbf{5}, 87 (2019).

\bibitem{Dehollain2020}
J. P. Dehollain, U. Mukhopadhyay, V. P. Michal, Y. Wang, B. Wunsch, C. Reichl, W. Wegscheider, M. S. Rudner, E. Demler, and L. M. K. Vandersypen,
Nature \textbf{579}, 528 (2020). 

\end{thebibliography}
\end{document}